\documentclass[
10pt
,reprint
,superscriptaddress
,twocolumn
,showpacs
]{revtex4}%

\usepackage{amsfonts}
\usepackage{amsmath}
\usepackage{amssymb}
\usepackage{array}
\usepackage{graphicx}
\usepackage{amsmath}
\usepackage{amssymb}
\usepackage{stmaryrd}
\usepackage{url}
\usepackage{natbib}

\newcommand{\unit}[1]{\ensuremath{\,\mathrm{#1}}}
\newcommand{\mgf}{\ensuremath{\mathrm{MgF}_2}}

\newcommand{\mvec}[1]{\ensuremath{\vec {#1}}}

\begin{document}

\title{Thermal-noise limited laser stabilization to a crystalline whispering-gallery-mode resonator}

\author{J.~Alnis}
\affiliation{Max-Planck-Institut f{\"u}r Quantenoptik, 85748 Garching, Germany}
\author{A.~Schliesser}
\email{albert.schliesser@mpq.mpg.de}
\affiliation{Max-Planck-Institut f{\"u}r Quantenoptik, 85748 Garching, Germany}
\affiliation{Ecole Polytechnique F$\acute{e}$d$\acute{e}$rale de Lausanne (EPFL), CH-1015 Lausanne, Switzerland}
\author{C.~Y.~Wang}
\affiliation{Max-Planck-Institut f{\"u}r Quantenoptik, 85748 Garching, Germany}
\author{J.~Hofer}
\altaffiliation[Present address: ]{Paul-Scherrer-Institut, CH-5232 Villigen, Switzerland}
\affiliation{Max-Planck-Institut f{\"u}r Quantenoptik, 85748 Garching, Germany}
\author{T.~J.~Kippenberg}
\affiliation{Max-Planck-Institut f{\"u}r Quantenoptik, 85748 Garching, Germany}
\affiliation{Ecole Polytechnique F$\acute{e}$d$\acute{e}$rale de Lausanne (EPFL), CH-1015 Lausanne, Switzerland}
\author{T.~W.~H{\"a}nsch}
\affiliation{Max-Planck-Institut f{\"u}r Quantenoptik, 85748 Garching, Germany}

\begin{abstract} 
We have stabilized an external cavity diode laser to a  whispering gallery mode resonator formed by a protrusion of a single-crystal magnesiumdifluoride cylinder.
The cylinder's compact dimensions ($\lesssim1\unit{cm}^3$) reduce the sensitivity to vibrations and simplify the stabilization of its temperature in a compact setup.
In a comparison to an ultrastable laser used for precision metrology we determine a minimum Allan deviation of $20\unit{Hz}$, corresponding to a relative Allan deviation of $6\times10^{-14}$, at an integration time of $100\unit{ms}$.
This level of instability is compatible with the limits imposed by fundamental fluctuations of the material's refractive index at room temperature. 
\end{abstract}


\maketitle

In many applications of lasers in metrology, sensing and spectroscopy, the quality of a measurement critically depends on the spectral purity of the employed laser source. 
For this reason, researchers' efforts have been directed towards the reduction of the laser's phase or frequency fluctuations since the early days of laser science \cite{White1965}.
Tremendous progress has been made since then, culminating in light sources with linewidths below $1\,\unit{Hz}$, and a relative Allan deviation of their frequency below the $10^{-15}$-level \cite{Young1999, Jiang2011}.
Such ultrastable lasers are operated in a number of specialized laboratories, and are realized by electronically locking the laser frequency to a resonance of an external reference cavity defined by two high-reflectivity mirrors attached to a low-thermal-expansion spacer  \cite{Stoehr2006, Ludlow2007, Webster2008, Alnis2008, Millo2009}.
Their performance has been enabled by sophisticated engineering of the  $\gtrsim1\unit{dm}^3$-sized spacers' material and geometry, as well as their vibration and thermal isolation. 
It is now understood to be limited by thermodynamic fluctuations in the mirror substrates and coatings \cite{Numata2004, Notcutt2006}.
Several proposals to overcome this limitation are discussed in the literature, including cryogenic operation of reference cavities  \cite{Kimble2008a, Gorodetsky2008, Meiser2009, Seel1997}.

\begin{figure}[tbp]
\begin{center}
{\includegraphics[width=\linewidth]{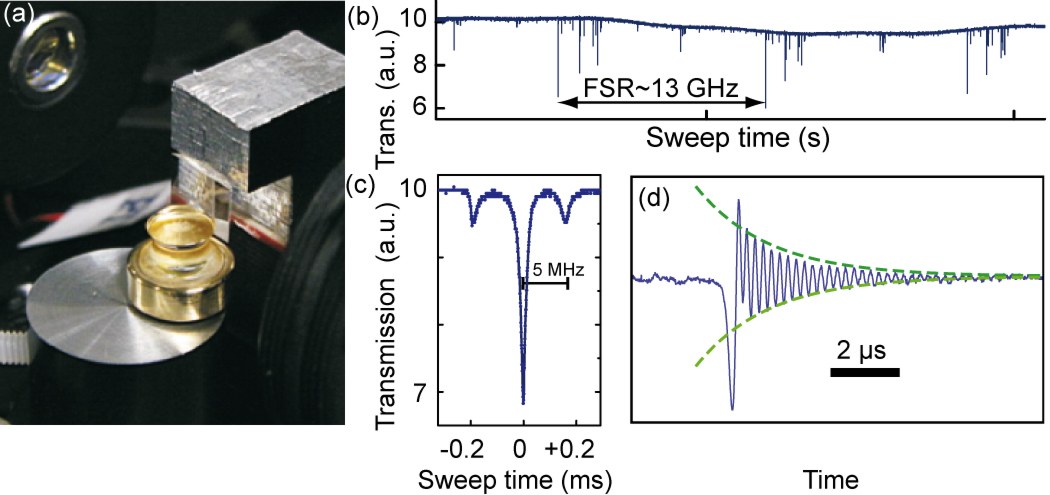}  }
\end{center}
\caption{
$\mathrm{MgF}_2$ WGM resonator (diameter $5\,\unit{mm}$) used for laser frequency stabilization. (a) Photograph of the resonator and the coupling prism in a simple coupling setup. (b) Transmission for a laser scan over more than 3 free spectral ranges showing different transverse WGMs.  (c) Typically observed  WGM resonance with laser-limited $0.6\,\unit{MHz}$ linewidth (sidebands are due to laser frequency modulation at $5\unit{MHz}$). (d) Ringing in the transmission signal observed in a fast laser sweep. The envelope of the oscillations (dashed green line) indicates an exponential field decay from the WGM within $\tau\approx2{.}1\unit{\mu s}$, corresponding to $Q\approx 2.0\times10^9$. 
}%
\label{f:WGM}
\end{figure}

In this work, we explore a different approach \cite{Matsko2007, Savchenkov2007a} to provide a reference for the laser's frequency, by fabricating whispering-gallery-mode (WGM) resonators from a single crystal of magnesium fluoride (Fig.~\ref{f:WGM}).
The WGM resides in a protrusion of a \mgf\ cylinder, whose compact dimensions ($\lesssim 1\,\unit{cm}^3$) and monolithic nature inherently reduce the resonator's sensitivity to vibrations.
It also enables the operation in more noisy and/or space-constrained environments, such as a cryostat or a satellite.
Furthermore, in contrast to the highly wavelength-selective and complex multilayer coatings required for mirror-based resonators, WGM resonators are intrinsically broadband, limited only by optical absorption in the host material.

Experiments with silica microspheres \cite{Vassiliev1998, Kieu2007, Sprenger2009} have already demonstrated the potential to realize compact, narrow-linewidth laser sources with WGM resonators.
But only in recent experiments  \cite{Sprenger2010, Liang2010} with crystalline $\mathrm{CaF}_2$ WGM resonators has the laser frequency  stability  been quantitatively assessed with sufficient resolution.
The lowest relative Allan deviations have been determined at the $10^{-12}$-level for sub-millisecond integration times.
We are able to significantly improve this performance and explore the thermodynamic limits of this approach at room temperature.

We fabricate WGM resonators following the pioneering work of Maleki and co-workers \cite{Grudinin2006,Grudinin2006a}, combining a shaping and several polishing steps on a home-built precision lathe \cite{Hofer2009}.
Several \mgf\ resonators were produced with a typical radius of $2~\unit{mm}$, one of which is shown in Fig.~\ref{f:WGM}a).
The WGMs are located in the rim of the structure, which has been fine-polished with a diamond slurry of $25~\unit{nm}$ average grit size in the last step.
The resulting surface smoothness, together with very low absorption losses in the ultra-pure crystalline material (Corning), enables quality factors in excess of $2\times10^9$.

A high index-prism (GGG, gadolinium gallium garnet) is used to couple the beam of an external cavity diode laser into the WGM.
Figure \ref{f:WGM} shows the transmitted power, i.~e.\ the optical power detected at the output of the coupling prism.
For laser stabilization, we choose to work in the undercoupled regime,
in which the presence of the coupling prism has only a negligible effect on the WGM quality factor $Q$ and resonance frequency $\omega_\mathrm{c}$ (see below).
To accurately determine the linewidth $\kappa\equiv \omega_\mathrm{l}/Q$ of such a resonance, we sweep the laser frequency $\omega_\mathrm{l}=2\pi c/\lambda$ ($c$ is the vacuum light speed, $\lambda= 972\unit{nm}$ the laser wavelength) through the WGM resonance at a rate greater than $\kappa^2$.
When the circulating power built up in the WGM decays during the ringdown time $\kappa^{-1}$, it interferes with the laser field swept already beyond the cavity resonance.
A transient, chirped heterodyne beat is therefore observed in the transmitted power (Fig.~\ref{f:WGM}c), the envelope of which decays with the cavity field \emph{amplitude} ringdown time $2\kappa^{-1}$.
The linewidth determined in this way is unobscured by the linewidth of the probing laser, and is given by  $\kappa=2\pi\times150~\unit{kHz}$ for the WGM used in this work (finesse $\mathcal{F}=87{,}000$).

\begin{figure}[tbh]
\begin{center}
{\includegraphics[width=\linewidth]{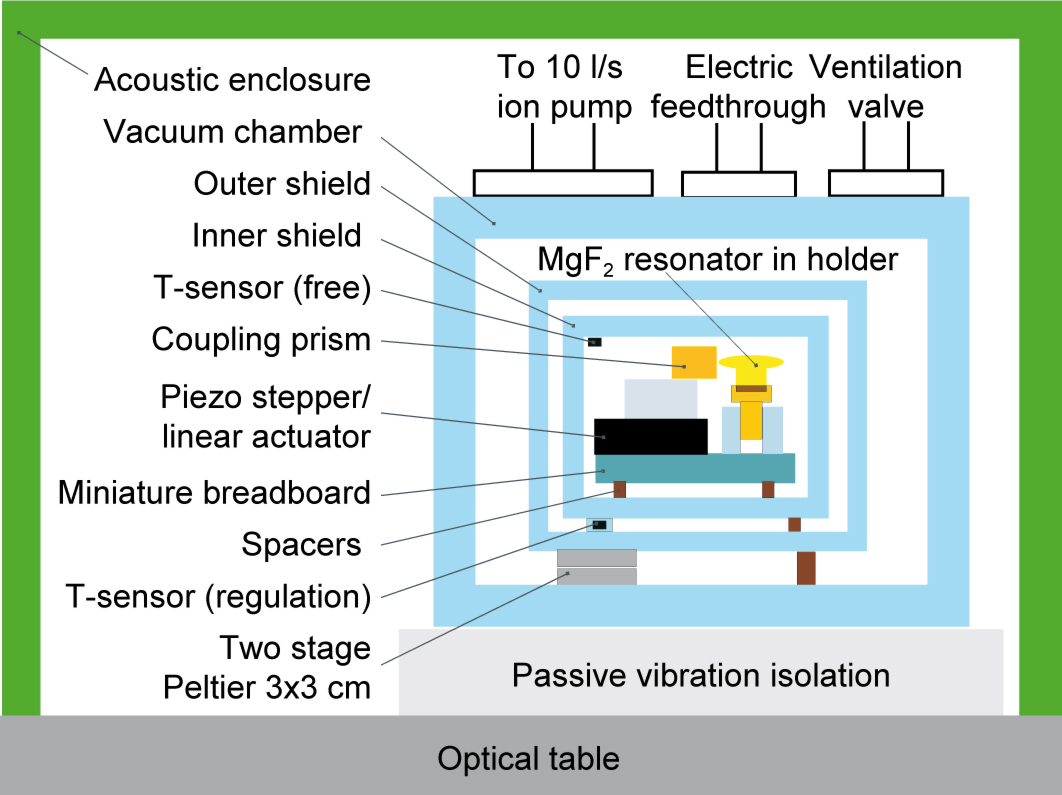}  }
\end{center}
\caption{
Thermal and acoustic isolation of \mgf\ cavity and coupling setup.
See text for detailed description.}
\label{f:isolation}
\end{figure}

For the laser stabilization experiments, a $\mathrm{MgF}_2$ resonator is mounted into a prism-coupling setup shielded against vibrations and thermal fluctuations.
In this proof-of concept experiment (Fig.~\ref{f:isolation}), we have made no attempts at miniaturizing the assembly, but have drawn on standard techniques developed for isolating larger reference resonators from their environment \cite{Alnis2008}: 
To suppress acoustic perturbations, the setup is kept in high vacuum ($p<10^{-6}~\unit{mbar}$) maintained in a $15\times17\times12~\unit{cm}^3$ aluminum chamber by an ion pump.
An additional box of heavy loudspeaker plywood shields the experimental setup from acoustic noise, while a passive vibration platform (Minus-K) and a standard optical table suppress the transmission of low-frequency vibrations of the laboratory environment to the coupling setup.

The resonance frequency $\omega_\mathrm{c}$ of the WGM sensitively depends on the temperature of the resonator due 
to induced changes in the radius $R$ and refractive index $n$ of the device, as described by the respective coefficients $\alpha_\mathrm{l}^\mathrm{}=\frac{1}{R}\frac{\mathrm{d}R}{\mathrm{d}T}$ and $\alpha_\mathrm{n}=\frac{1}{n}\frac{\mathrm{d}n}{\mathrm{d}T}$.
Since $\alpha_\mathrm{l}$ is as large as $10^{-5}/\mathrm{K}$ in \mgf\ at room temperature, it is important to decouple the resonator from the unstable ($\pm 0.2~\unit{K}$) laboratory temperature.
The low pressure in the chamber already suppresses heat exchange through the residual gas, while heat conduction through the support of the coupling setup is reduced by minimizing the contact area to three thermally insulating spacers.
Heat exchange via thermal radiation is reduced by introducing two additional aluminum shields between the chamber's walls and the actual coupling setup (Fig.~\ref{f:isolation}).
These measures increase the inertial time for the transmission of temperature changes from the laboratory environment to the resonator to several hours.
In addition, electronic feedback to a two-stage thermoelectric element is used to actively stabilize the temperature of the outer shield to a value
close to room temperature.
With the loop activated, the inner shield's temperature is stable to $1\unit{mK/day}$
according to an out-of-loop temperature sensor.

\begin{figure}[tbhp]
\begin{center}
{\includegraphics[width=.841\linewidth]{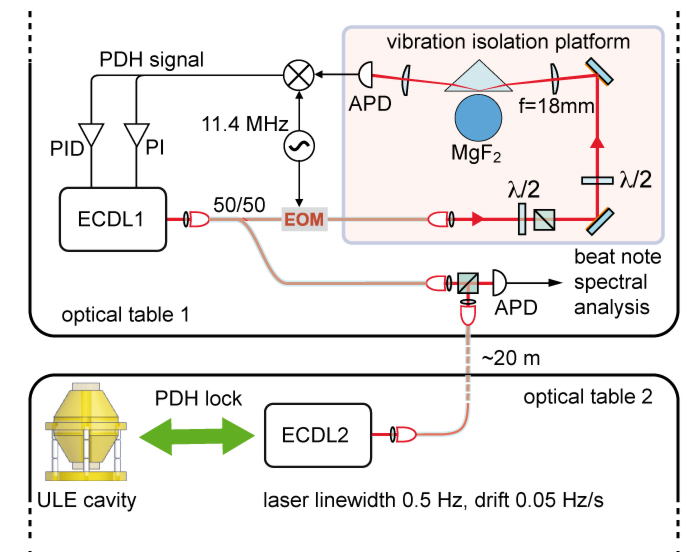}  }
\end{center}
\caption{
Laser stabilization to a WGM in a \mgf\ resonator using the Pound-Drever-Hall (PDH) method, and comparison to an ultrastable laser locked to a state-of the art reference cavity on a second optical table.
ECDL: external cavity diode laser 1/2;
EOM: electro-optic modulator;
$\lambda/2$: half-wavelength retardation plate;
APD: avalanche photodiode;
PI(D): proportional-integral-(differential) feedback controller;
ULE: ultra-low expansion glass.
}%
\label{f:setupAlbert}
\end{figure}

The beam of a commercial Littman-type extended cavity diode laser (Velocity, New Focus) is focused on the face of the coupling prism through anti-reflection coated windows in the vacuum chamber and small (diameter $1~\unit{cm}$) bore holes in the aluminum shields.
The laser is locked to a high-$Q$ WGM using the Pound-Drever-Hall method \cite{Drever1983}
implemented here with  an external electro-optic phase modulator driven at $11{.}4~\unit{MHz}$ and demodulation of the transmission signal at the same frequency (Fig.~\ref{f:setupAlbert}).
The power, polarization, position and pointing of the laser beam incident on the coupling prism, as well as the prism-resonator gap are then optimized for a maximum error signal.
The resulting error signal is fed back via a two-branch control system actuating both the grating tilt in the laser (via a piezoelectric transducer) and the diode pump current.
The error signal is amplified by home-built proportional-integral (PI) controllers in both branches; the current controller has an additional phase advance (PID) for faster regulation up to a feedback frequency of ca.~$1~\unit{MHz}$.

\begin{figure}[bth]
\begin{center}
{\includegraphics[width=\linewidth]{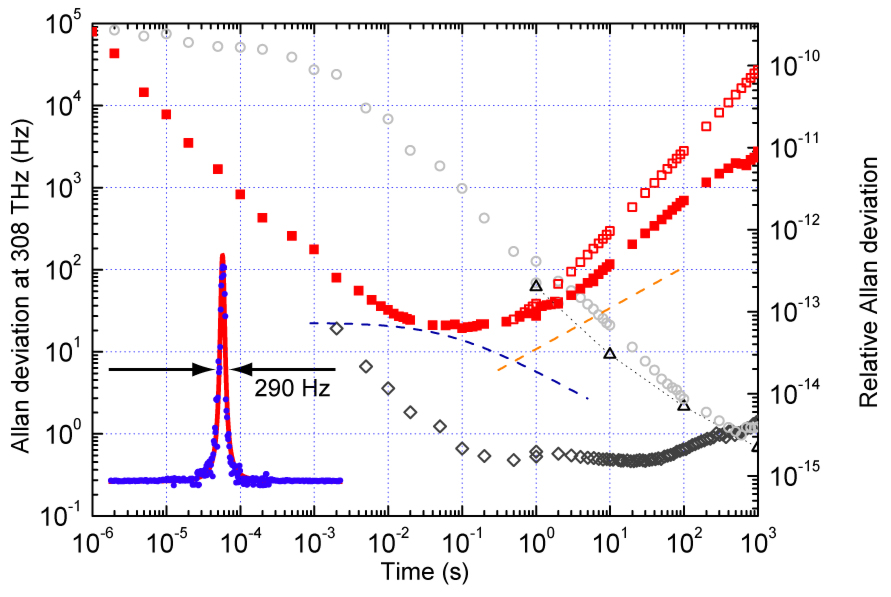}  }
\end{center}
\caption{
Allan deviation of the optical beat note in Hz, and normalized to the optical carrier at 308 THz.
Inset shows a raw beat note measured  at $586\unit{MHz}$ with a spectrum analyzer, with an analyzer-limited width of $290\unit{Hz}$.
Diamonds are reference measurement of beat note stability between two lasers locked to two mirror-based resonators \cite{Alnis2008}.
Red squares are Allan deviation of beat frequency between two lasers locked to a mirror-based and the WGM resonator with a linear drift of $38 \unit{Hz/s}$ removed. Open squares show data without removing drift.
For comparison, gray circles show the Allan deviation measured between a laser stabilized to a two-mirror cavity and one tooth of an optical frequency comb stabilized by a hydrogen maser, the specified relative Allan deviation of which is indicated by triangles.
Blue and orange dashed lines are the estimated minimum Allan deviation due to thermorefractive and photothermal noise, respectively.
}
\label{f:allan}
\end{figure}

To assess the frequency fluctuations of the laser locked to the WGM resonator, its frequency is compared to the one of an independent diode laser locked to an ultrastable mirror-based resonator in the same laboratory \cite{Alnis2008}.
The extraordinary stability of the latter affords a direct measurement of the WGM-stabilized laser by analyzing the spectral properties of the radio-frequency beat generated between the two lasers in a heterodyne detector.
Figure \ref{f:allan} (inset) shows the radio-frequency spectrum of the beat recorded at $586\unit{MHz}$.
Its width was fitted to  $290~\unit{Hz}$, limited by the resolution of the spectrum analyzer.
For a more systematic characterizations of the laser's stability, we determine the Allan deviation $\sigma_y(\tau)$ of the beat's frequency as a function of the gate time $\tau$ of the electronic frequency counters employed to that end.
The results of these measurements are shown in Fig.~\ref{f:allan} along with data from earlier experiments \cite{Alnis2008}.
On short ($\ll 1~\unit{ms}$) time scales, the measured Allan deviation is inversely proportional to the gate time, as expected for white phase noise.
We attribute this noise to uncompensated fluctuations to the diode lasers' phase due to the limited feedback bandwidth and gain of the stabilization electronics.
For the same reason, we also expect Brownian noise in the cavity due to thermal excitation of its mechanical modes---as observed recently in  CaF$_2$ WGM resonators \cite{Hofer2009}---to be insignificant in this measurement.

For gate times between ca.~$10~\unit{ms}$ and $1~\unit{s}$, the Allan deviation is approximately constant at a level of several $10~\unit{Hz}$, corresponding to a relative Allan deviation below \mbox{$10^{-13}$} (Fig.~\ref{f:allan}).
For gate times beyond $1~\unit{s}$, the Allan deviation increases again, with an asymptotic $\sigma_y(\tau)\propto \tau$ dependence for long gate times, indicative of linear frequency drift.
The extracted drift rate of $38\unit{Hz/s}$ is compatible with drift of the resonator temperature due to the imperfect stabilization of the latter and can be subtracted from the data for further analysis. 
The residual Allan deviation at longer time scales ($\tau>10\unit{s}$) is equally expected to be caused mainly by temperature instability.
Other sources specific to this setup---such as the resonator-prism gap controlled by a piezoelectric transducer---were checked to be uncritical ($|\partial \omega_\mathrm{c}/\partial U|\approx2\pi\, 35\unit{Hz/mV}$, where $U$ is the piezo voltage).
Importantly, temperature fluctuations can not only be induced by a changing environment temperature, but also by fluctuations of the absorbed optical power, as caused by amplitude fluctuations of the intracavity laser field.
Indeed, the large measured resonance frequency dependence on the power sent into the coupling setup ($\partial \omega_\mathrm{c}/\partial P_\mathrm{in}\approx- 2\pi\, 28\unit{kHz/\mu W}$) indicates that \emph{photothermal noise} could be another important source of instability.
In an independent measurement, we have determined the spectrum $S_{PP}(\Omega)$ of the power fluctuations of the diode laser in the (Fourier) frequency range between $30\unit{mHz}$ and 
$300\unit{Hz}$.
The measured relative fluctuations of $\lesssim6\times 10^{-5}$ in this low-frequency band correspond to fluctuations of the cavity frequency by $25\unit{Hz}$ for a launched power of $15\unit{\mu W}$.
Figure \ref{f:allan} shows the expected contribution to the Allan deviation assuming an instantaneous response of the cavity resonance frequency $S_{\omega\omega}(\Omega)=(\partial \omega_\mathrm{c}/\partial P_\mathrm{in})^2 S_{PP}(\Omega)$.

While the influence of environment temperature fluctuations and photothermal noise can be further suppressed by technical improvements, more fundamental sources of noise were observed to limit the lowest Allan deviations measured.
These noise sources originate from the fact that the temperature of a body of heat capacity $C_V$ at an average temperature $T$ intrinsically fluctuates as $ \langle \delta T^2Ê\rangle= {k_\mathrm{B} T^2}/{C_V}$, where $k_\mathrm{B}$ is Boltzmann's constant \cite{Landau1980}.
This implies, for example, that the temperature of the resonator fluctuates by $\langle\delta T_\mathrm{r}^2\rangle=k_\mathrm{B} T^2 / c_V \rho V_\mathrm{r}\approx (2\unit {nK})^2$ with the specific heat capacity $c_V=1020\unit{J/kg\,K}$ and density $\rho=3180\unit{kg/m^3}$ of \mgf \cite{Weber2003}.
The resulting fluctuations of the resonator radius \cite{Matsko2007} lead to comparably small resonance frequency fluctuations on the order of $\langle\delta \omega_\mathrm{c}^2\rangle=\omega_\mathrm{c}^2  \alpha_\mathrm{l}^2\langle\delta T_\mathrm{r}^2\rangle\approx(2 \pi \, 7\unit{Hz})^2$.
In contrast, the fluctuations of temperature $T_\mathrm{m}$ sampled by the optical mode are larger due to the significantly smaller mode volume  \cite{Gorodetsky2004} $V_\mathrm{m}={(\int |\mvec{E}|^2 \mathrm{d}^3 r)^2}/{\int |\mvec{E}|^4 \mathrm{d}^3 r}$ of approximately $2\times10^{-12}\unit{m^3} $ in the used resonator, leading to $\langle \delta T_\mathrm{m}^2\rangle=(0.4 \unit{\mu K})^2$.
This fluctuating temperature directly modulates the refractive index in the WGM and, as a consequence, its resonance frequency according to $\langle\delta \omega_\mathrm{c}^2\rangle=\omega_\mathrm{c}^2  \alpha_{n}^2\langle\delta T_\mathrm{m}^2\rangle\approx(2 \pi \, 60\unit{Hz})^2$, where we have used the mean value of $\alpha_n=5\times 10^{-7}/\mathrm{K}$ for ordinarily and extraordinarily polarized light at 972~nm.
The large fluctuations due to this \emph{thermorefractive noise} \cite{Braginsky2000,Gorodetsky2004} therefore  limit the frequency stability of the WGM.

For a quantitative comparison with our measurements, we calculate the expected Allan deviation as a function of measurement time.
Approximating the resonator as an infinitely long cylinder of \mgf, the (double-sided) spectral density of fluctuations of the average temperature $\bar u$ of the mode can be estimated to \cite{Braginsky2000, Gorodetsky2004,Matsko2007}
\begin{equation}
  S_{\bar uÊ\bar u}(\Omega)\approx\frac{ k_\mathrm{B} T^2}{c_V \rho V_\mathrm{m}} \sum_p 
  \int
  \frac{2 \pi D R\, \Lambda^2(k_p,k_m)}{\Omega^2+D^2 k^4} \frac{\mathrm{d}k_m}{2\pi},
  \label{e:suu}
\end{equation}
where $k^2=k_m^2+k_p^2$, and $D=\lambda^*/c_V \rho\approx 7.9\times 10^{-6}\unit{m^2/s}$ is \mgf's diffusivity (assuming, for simplicity, a mean thermal conductivity of $\lambda^*\approx26\unit{W/K\, m}$).
The wavenumbers $k_p$ ($k_m$) characterize the thermal waves in radial (axial) direction permitted by the boundary conditions, implying in particular that $k_p R$ are the roots of the first-order cylindrical Bessel function in this model.
The overlap function $\Lambda(k_p,k_m)$ of the thermal and optical modes is approximately constant for all thermal waves with spatial frequencies $k/2\pi$ that are much smaller than the inverse transverse dimension $b^{-1}$ of the optical mode.
Thermal modes with larger spatial frequencies make only small contributions to the sum/integrals at Fourier frequencies $\Omega\ll D b^{-2}$, so that the spectrum of WGM frequency fluctuations can be calculated from eq.\ (\ref{e:suu}) using
  $S_{\omega\omega}(\Omega)=\omega_c^2 \alpha_n^2 S_{\bar uÊ\bar u}(\Omega)$ with 
  $\Lambda(k_p,k_m)\approx 1/\sqrt{\pi}R $.
Figure 4 shows the resulting frequency fluctuations converted to an Allan deviation.	
In spite of this only crude estimation, it is clear that thermorefractive noise constitutes a critical limitation for WGM frequency references.
The lowest Allan variances determined in our measurements are more than one order of magnitude above the level expected from laser phase noise and thermal drifts, and compatible with estimations for thermorefractive noise.

In conclusion, we have demonstrated the stabilization of a diode laser by locking it to a crystalline WGM resonator used as an optical reference, and characterized the stability of the laser's frequency by comparison with an ultrastable optical reference. 
At an integration time of $100\unit{ms}$, the lowest measured Allan deviation was $20 \unit{Hz}$, a value which is compatible with thermorefractive fluctuations in the resonator material.
While the stability does not yet reach the level achieved by optical references based on two-mirror resonators, WGM resonators could be used to significantly reduce the phase noise of diode lasers or free-running frequency combs in simple, compact setups.
The references' performance is expected to improve if the strong temperature sensitivity can be reduced or eliminated.
For example, the operation close to temperatures ($\sim200^\circ \unit{C}$) at which the thermorefractive coefficients $\alpha_n$ of \mgf\ vanish \cite{Savchenkov2007a} is practically feasible.
On the other hand, self-referenced temperature stabilization \cite{Matsko2011} could dramatically improve temperature stability.
Finally, the inherent compatibility of these resonators with cryogenic operation opens a promising approach not only to an improved temperature stability and reduced sensitivity to temperature fluctuations, but also a strong suppression of thermodynamic fluctuations limiting also today's best optical flywheels. 

\begin{acknowledgments} 
The authors would like to thank M.\ Gorodetsky for valuable discussions.
 TJK acknowledges funding by an Marie Curie IAPP and the Swiss National Science Foundation.
\end{acknowledgments}

\end{document}